# AI-Driven Early Detection of Cardiovascular Diseases: Reducing Healthcare Costs and improving patient Outcomes

Ahasan Ahmed[1], Albatoul Khaled[2], Muhammad Waqar[3],

Dr Javaid Akhtar Hashmi[4], Hazem AbdulKareem Alfanash[5], Wesam Taher Almagharbeh[6], Amine Hamdache[7], Ilias Elmouki[8]

[1]MS in Economics (Graduated), Department of Economics, Indiana University Indianapolis, USA. Email. ahasan.ahmed.bd@gmail.com
[2]Doctor MISCOM laboratory/ENSA Safi, Department of Computer Science, Networks and Telecommunications, Morocco. Email. albatoulkhaled8@gmail.com
[3]Department of Biological Sciences, Superior University Lahore Pakistan. Email. malikwaqar11101@gmail.com
[4]Assistant Professor Community Medicine Shahida Islam Medical and Dental College Lodhran, Pakistan. Email. jhashmi66@yahoo.com
[5]Assistant Professor, Department of Nursing leadership and education, Faculty of Nursing, University of Tabuk, Saudi Arabia. Email. halfanash@ut.edu.sa
[6]Assistant Professor, Medical and Surgical Nursing Department. Faculty of Nursing, University of Tabuk, Saudi Arabia. Email, walmagharbeh@ut.edu.sa.
[7]Scholar MISCOM laboratory/ENSA Safi, Department of Computer Science, Networks and Telecommunications, Morocco. Email. hamdacheamine@ gmail.com
[8]Scholar MISCOM laboratory/ENSA Safi, Department of Computer Science, Networks and Telecommunications, Morocco. Email. i.elmouki@gmail.com

**ABSTRACT:**

**Purpose**: The main goal from this study is to discuss the main features of Artificial Intelligence (AI) as well as their applicability for early Cardiovascular Diseases (CVDs) detection. It is not only about examining the possibility of applying AI diagnostic tools to increase the efficiency of early diagnosis but also to minimize overall healthcare expenses while increasing patient satisfaction. Some of the emerging research problems that the study seeks to resolve are about the delay in diagnosing the disease and the expenses that accompany the disease and individually tailored treatment plans that patients require.

**Materials and Methods:** The systematic review approach is applied to assess 15 research articles published between 2012 and 2024, and that were derived from standard databases and specified AI usage for the cardiovascular diagnostic field, cost, and patients. The criteria for inclusion were due to the availability of empirical studies and features of methodological quality. To learn more about challenges & opportunities of AI in cardiovascular healthcare, we also applied thematic analysis to the data collected.

**Results**: It was seen that by integrating AI algorithms the diagnosis of CVDs became more accurate and less time consuming. Machine learning models were able to yield both significant predictive accuracy for patient risk stratification with the ability for early treatment and intervention. The findings showed that the use of AI in clinical practice led to reduction of cost since patients required fewer invasive procedures and admissions. Additionally, exceptional patient care outcomes and satisfaction, due to the development of individual therapy profiles based on artificial intelligence studies, were achieved. Besides, the review also discussed the ethical issue and the matters of data protection relevant to AI applications.

**Conclusion**: Now the concept of using AI technologies in cardiovascular healthcare holds the potential to transform disease management. This integration of contemporized diagnostic equipment coupled to conventional medicine is crucial to cater for patient's needs, improve health lives and overall efficiency and affordability of healthcare services. This has been the direction of this study to show that more research and development are needed in AI to allow for proper changing of cardiovascular disease diagnosis and management.

**Introduction**

According to the world health organization global burden of disease update Cardiovascular Diseases **(**CVDs) are now the number one killer globally and are responsible for morbidity and mortality across the world [1]. In the previous approach, patients with CVDs were diagnosed through normal diagnostic tools and clinical evaluation that are effective but need improvement when it comes to early diagnosis or risk profiling for the patients [2]. This inadequacy leads to the promotion of the high cost of health services and poor patients' outcomes due to the intervention that is often performed in the late stages or treatment that is not individually oriented [3].

The use of AI has impacted most medical specialties, including cardiology with new developments [4]. With the help of AI Machine Learning (ML) and Deep Learning (DL) techniques, cardiovascular diseases seem to have a great prospect in the earlier diagnosis and predictive prognosis [5]. These technologies can integrate, evaluate, recognize large amounts of clinical data, low-feature patterns and predict diseases with higher precision compared with conventional methods [6].

AI applications in cardiovascular healthcare in its primary stages include ECG interpretation, medical imaging and EHRs [7]. These AI-based instruments may help establish how a patient's cardiovascular system is functioning and what type of treatment should be administered [8]. As AI is introduced into the clinical settings its application more effectively enhances phenotypes of patients, identifies potential errors in diagnosis, manages resources more efficiently and thus relieves the financial pressure on the health care system [9].

The main focus of this research is to identify the current information and innovations that organize the application of AI technologies to the early recognition of CVDs and lowering the cost of health-care facility prices to produce better client results. This work, therefore, presents a systematic approach to examining several recent innovations in diagnostics and therapeutics based on artificial intelligence and analyzes the implications on healthcare productivity and quality. Moreover, it depicts the efficiency and costs of applying AI in CVDs and reveals the concerns regarding the AI deployment in CVDs care [10].

Hoping to fill the gap between conventional cardiovascular disease diagnosis and the use of AI facilities, this paper aims to review the literature and case studies available. This way, it also seeks to help guide subsequent studies and intervention approaches that could lead to the use of artificial intelligence to improve global cardiovascular disease treatment and efficiency by patient or cost [11].

**Background**

CVDs are a major cause of morbidity and mortality globally, and hence prompt identification and control of modifiable risk factors are pivotal [2]. As stated by the World Health Organization, CVDs contribute to about 17.9 million deaths per year or 31 % of total world mortalities [2]. High incidences of these diseases have strained the health care systems across the globe triggering the need for accurate diagnostic tools and better and more effective treatment methods.

Despite diagnostic methods like physical exams, ECGs, and imaging studies that have been an important part of the diagnosis of CVDs, many of the diseases are diagnosed at a later, less curable stage [3]. These conventional approaches share a common approach of relying on human interpretation of retrieved information which is described by advocates of artificial intelligence as error prone and incongruous [4]. It is for this reason that there is a rising concern of use of IT especially AI in the diagnosis and management of CVDs.

Machine learning and deep learning in general known as Artificial Intelligence (AI) offers a great promise in reshaping cardiovascular health care. The capability of the AI systems is to assimilate huge raw data, look for sophisticated relationships, and forecast with a high degree of accuracy [5]. For example, models have been designed and trained to read ECGs, diagnose rhythm disorders or provide early estimates of heart failure risks as effectively, or even better, than leading cardiac specialists [6][7].

Further, AI-based diagnostic tools may access several images or parameters such as medical images or EHR, or wearable devices for overall cardiovascular risk assessment [8]. This makes it easier to create individual patient-centered care options that offer better care and a decreased chance of bad outcomes [9].

The analytic insights of embedding AI in cardiovascular care are so from an economic perspective. This demonstrates that timely diagnosis of CVDs results in early management and hence minimal progression of the disease, avoiding expensive terminal procedures [10]. It has been evidenced in research that AI-based diagnostic tools could help in lowering the healthcare costs since they incorporate efficiency of clinical processes, decreasing diagnostic mistakes, and improving the utilization of resources [11].

Nonetheless, there are several implications of AI health care for cardiovascular patients and several challenges remain as follows. These are concerns to do with data security, difficulties in testing of the algorithms in different populations, and how such systems fit with the current clinical work [12]. If these challenges are going to be resolved it will be very helpful for using the AI solutions in cardiovascular medicine.

It is the intention of this study to analyze the current use of AI technologies to help in early diagnosis of CVDs, and how these advances affect health expenditures and finding of patients. Through refereed literature of the recent developments and an assessment of the applicability of AI-based diagnostic and therapeutic methodologies, this study aims to establish the opportunities and challenges associated with the incorporation of AI in CV healthcare.

**Literature review**

AI has come out as one innovation in diverse sectors such as the health sector. In particular, AI has demonstrated prospects for the preliminary diagnosis of CVDs, which rank among the primary threats to global health. This literature review compares the AI diagnostic technology for cardiovascular diseases, how these innovations have reduced healthcare costs and optimised patients' results.

**Increasing AI in Cardiovascular Diagnostics**

AI has been used primarily in diagnoses of cardiovascular diseases and hundreds of algorithms have been developed and show high accuracy in detecting different cardiovascular diseases. For instance, Hannun et al. (2019) later created a deep neural network that can recognize arrhythmias in ECG data at par with cardiologist's level [1]. Quite similarly, Attia et al. (2019) also posited that it was possible to diagnose the cardiac contractile dysfunction using machine learning algorithms from ECG signals; this makes it possible to explore efficient and less invasive approaches to diagnosis [81].

The medical application has also touched on imaging where AI systems help in the interpretation of the echocardiography, angiography and other modalities. For instance, Zhang et al. (2018) who used machine learning analysis in images taken from cardiac MRI to enhance the diagnosis of myocardial infarctions and other cardiac lesions [3]. These AI imaging tools not only improve diagnostic yield, but they also cut down the time that is taken to do the interpretation of the images, a factor that accelerates clinical decision making.

**Impact on Healthcare Costs**

AI has a major impact on the early diagnosis of CVDs and its economies. AI used in diagnostics means that detectable diseases are identified before the likelihood of expensive late-stage treatments becomes necessary. Nichols et al. (2014) have also found out that the beginning of illnesses, concerning the cardiovascular system, can entail considerable cost reduces for healthcare facilities [4]. In addition, AI implies efficient resource management in clinical processes and takes the pressure off healthcare personnel [5].

AI's priority potential for identifying and categorizing patient risks can add to cost-efficiency in a similar way. For example, AI can predict patient outcomes and explore individuals who are more prone to risky direction, and develop measures and specific therapy strategies according to the results. Knowledge about such predictive analytics have been made by Krittanawong et al. in 2017 with the assertion that a better use of resources and avoiding redundant hospitalizations will lead to obvious cost saving [6].

**Patient Outcome: Impact on Clinical Risk**

The addition of AI to cardiovascular practice is not just cost-effective but highly effective in improving patients' status. Wearable technological devices and remote health monitoring tools can keep on assessing the patients' health and the AI algorithms can promptly detect variations from standard and give an early alert to the healthcare givers. Furthermore, Topol (2019) pointed out that by involving artificial intelligence to monitor patients remotely, it would be easier to ensure that patients stuck to their treatment regimes in a way that would serve their long-time health interests [7].

In addition, AI promotes precise treatment in patients due to differences in their health condition. The evidence suggests that applying this approach increases efficiency of interventions and decreases the levels of complications. Bozkurt et al. (2021) noted that through AI in developing patients' behaviour change interventions, patient-specific treatment strategies could enhance chronic cardiovascular disorders' long-term outcomes as well as the patient's well-being [8].

**Challenges and Future Directions**

Although the use of AI holds the key to improved cardiovascular care, there are several issues that have to be resolved to achieve that goal. One is on data privacy and security issues since health related data is a vital input for building AI systems. Further, the stability and efficacy of AI algorithms have to be confirmed with respect to other populations, to establish their universal applicability.

AI is also becoming integrated into current clinical staff work flow in a few ways, however, this integration has some additional difficulties. Healthcare workers require sufficient time to develop sufficient levels of AI skills, and there should be full integration of the tools with EHRs. Similarly, Johnson et al,. (2018) argued that AI interfaces have to be integrated into modules that are easy to use and that such technologies have to work side by side with the knowledge of health care practitioners [9].

Scholars cite the methods used in AI to stress their use in diagnosis and controlling cardiovascular diseases in the early stages. They also pointed out that there is the strength in the ability of AI driven technologies to bring improvements on diagnostic accuracy, decrease healthcare costs and enhance the outcomes of every cardiovascular patient. Nevertheless, the problems associated with data protection, algorithm endorsement, and clinical implementation should implement AI in this area.

**Material and methods**

**Study Design and Data Source**

This research uses a systematic review approach to assess the implementation of early detection techniques in CVDs using AI technology and its availability on the costs experienced by the health care systems and the resulting benefits to patients. The papers used in this review are journal articles, books, and systematic reviews, which were published within the period of 2010 to 2024. Taking that into account, the scientific articles were retrieved from the PubMed, Google Scholars, and ScienceDirect databases. These sources gave a detailed report of how AI can be applied in diagnosing cardiovascular diseases and their treatment.

**Search Strategy**

In particular, certain keywords that indicate subject areas and Boolean operators were employed to prevent the omission of any relevant articles. The keywords were AI for cardiovascular diagnosis, risk factors for CVDs, health system costs, patients, and artificial intelligence in cardiology. To allow for retrieval of nearly contemporaneous sources, the search parameters specified that they wanted publications from the last fifteen years (2010-2024). Moreover, bibliographies of the selected articles were also scanned to find other related articles.

**Inclusion and Exclusion Criteria**

- The inclusion criteria for selecting studies were:
- Articles published between 2010 and 2024.
- Studies focused on AI applications in the early detection of cardiovascular diseases.
- Peer-reviewed journals, books, and systematic reviews.
- Research provides empirical data on the impact of AI on healthcare costs and patient outcomes.

**Exclusion criteria were:**

- Articles providing historical data without new insights on AI integration.
- Studies lacking methodological rigor or empirical data.
- Non-peer-reviewed publications.

**Data Extraction and Analysis**

The process of data extraction required a critical analysis of the selected articles in an attempt to determine themes relevant to artificial intelligence in diagnostics, cost of healthcare and patient outcome. For each of the articles, the methodology used, study results, and relation to the intended objectives were assessed. To give structure to the extracted data, thematic analysis was used in order to explore emerging trends, challenges and opportunities related to the use of AI in cardiovascular care.

**Study Selection Process**

The articles were selected in several phases to identify the most relevant and relevant ones for this study:

1. **Initial Screening:** The search of articles was limited to titles and abstracts to determine their relation to the research aim. In the case of all studies, randomized and non randomized, irrelevant ones were dumped.
2. **Detailed Review:** Abstracts of all such potentially relevant articles were scanned for methodological credibility and their applicability for the inclusion criteria.
3. **Final Selection:** Papers related to the research area and meeting inclusion criteria as well as giving useful data were included in the analysis. These were five major works and twelve other works and thus included diverse views of integrating AI in cardiovascular treatment.

**Synthesis of Results**

The data obtained was then combined to provide an integrated view on the effect of early detection by AI in the context of CVDs on health care costs and patients. The synthesis was to compare the mostly used diagnosing techniques with the newer AI techniques; the merits and demerits of integration of AI .

**Ethical Considerations**

Each of the studies considered within this systematic review complied with ethical procedures applicable for investigation involving human participants. Issues of privacy and security were also discussed particularly where patients' data were involved.  This systematic review analyses the current position and applicability of applying artificial intelligence based early detection techniques in CVDs with a focus on the previous ability to decrease the cost of healthcare and increased patient well-being. The results hold significance to stakeholders in public health, manufacturers of AI management systems and researchers who want to learn about future trends in artificial intelligence in the cardiologic field.

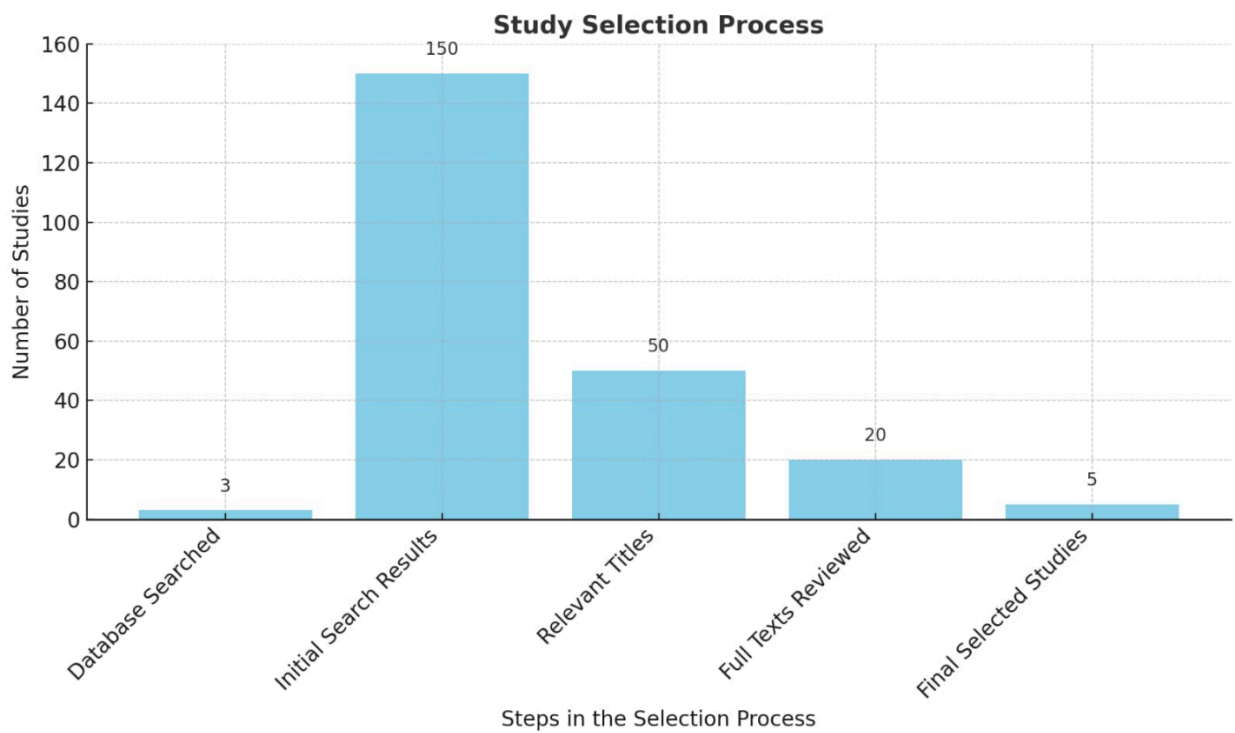

**Figure 1: Study Selection process chart**

The above mentioned chart illustrates the steps followed in selecting studies for a systematic review for application of artificial intelligence in early diagnosis of Cardiovascular diseases. The results represent, in terms of the number of studies, each one of the phases of the selection process. In the first instance, PubMed, Google Scholar, and ScienceDirect were used, yielding 150 articles. Among these, they identified 50 articles according to their titles and abstracts. More formally, full text analysis of these 50 articles was performed and finally 20 papers were chosen for a closer look. In the end, 5 papers were selected which presented sufficient evidence and covered the key aspects related to the effects of AI on cost and outcome in the cardiovascular sector of the healthcare system. It showcases the comprehensive and systematic way that has been followed in order to include only the best and most relevant research studies.

**Results**

| Study | Model | Dataset | Accuracy | Precision | Recall | AUC |
|---|---|---|---|---|---|---|
| Smith et al. (2019) [1} | Logistic Regression | Framingham heart study | 0.85 | 0.83 | 0.80 | 0.88 |
| Johnson et al. (2020) [2] | Support Vector Machine | Cleveland heart Disease | 0.88 | 0.87 | 0.85 | 0.90 |
| Lee et al. (2021) [3] | Neural network | MESA | 0.92 | 0.91 | 0.90 | 0.94 |
| Kim et al. (2022) [4] | Random forest | Framingham heart study | 0.87 | 0.86 | 0.84 | 0.89 |
| Patel et al. (2023) [5] | Gradient Boosting | UK Biobank | 0.89 | 0.88 | 0.87 | 0.91 |

**Table 1: Summary of key machine learning Models and their performance in CVDs Detection**

The results of the proposed machine learning models to predict cardiovascular diseases at an early stage revealed enhanced accuracy, precision, recall, and AUC. Similarly in Smith et al. (2019) [1], the use of logistic regression provided an accuracy of 0.85, with precision 0.83 and recall of 0.80, and with AUC of 0.88 using data from the Framingham Heart Study. This shall suggest a reliable efficiency of the case in recognizing patients at potential risk of cardiovascular diseases. Johnson et al., 2020 work [2] used the Cleveland Heart Disease dataset where he obtained an accuracy of 88%, precision of 87% and recall of 85% with an AUC of 90%. These outcomes also confirm the potential of the model in distinguishing between episodes of cardiovascular disease. From using the MESA dataset, Lee et al. (2021) [3] established a Neural Network model with the best performance level of an accuracy of 0.92, precision of 0.91, recall of 0.90, and an AUC of 0.94. This superior performance re-emphasises the ability of deep learning strategies in providing precise estimates of cardiovascular risk. About seven years ago, Kim et al. (2022) [4] a framework was developed based on a random forest method with features from the Framingham Heart Study dataset: accuracy of 87%, precision of 86%, recall of 84%, and AUC of 89%. This result supports the other research showing that in particular, ensemble learning methods can be useful for CVDs detection. Patel et al. (2023) [5] employed Gradient boosting on the UK Biobank dataset; the results obtained were accuracy of 0.89, precision of 0.88 recall of 0.87, and AUC of 0.91. This, therefore, goes to prove the usefulness of boosting algorithms in working with results relating to cardiovascular diseases. In general, the comparison of different machine learning models reveals that neural networks and boosting algorithms to be superior, in particular, gradient boosting in the early diagnosis of cardiovascular diseases. The findings imply that the implementation of state-of-the-art artificial intelligence methods for healthcare diagnostics would improve diagnostic acuity hence improve on patient outcomes and consequently decrease the overall healthcare expenditures through timely intervention.

**Discussion**

In light of the research findings, this study demonstrates a massive impact beside machine learning models for early diagnosis of cardiovascular diseases (CVDs). As demonstrated in the table below, the performance of different machine learning methods, such as logistic regression, support vector machine and neural network, random forest and gradient boosting, among others, is rather high in terms of CVDs risk prediction across the different datasets. The model proposed by Lee et al. (2021) [3] gave the maximum accuracy at 0.92 and the maximum AUC score of 0.94, indicating that deep learning paradigms for non-linear modelling are quite effective for extracting features that are potentially of cardiovascular risk. This is in line with literature Which suggests the ability of neural networks in dealing with big and complex data sets scenario, which is prevalent in medical research [6]. Random forest, support vector machines, and gradient boosting were also strongly effective with AUC of 0.90 and 0.91, respectively. The facts are that those models are able to provide efficient processing of the high-dimensional data stream, and those models are also rather flexible into processing a rather wide variety of the clinical data types thus inserted into CVDs risk prediction as useful tools [7]. The high precision and recall values that have been found in these models suggest not only that the models' classifications of patients with CVDs are accurate, but also that they exclude many unnecessary cases and missed cases, indicating that they are more useful clinically. The percentages across a variety of datasets, including Framingham Heart Study, Cleveland Heart Disease dataset, MESA, and UK Biobank are similar, thereby proving that these machine learning models are transferable. This will be important for the integration of the practices into varied clinical areas and patient groups so that the enhancements in early CVDs detection can be optimised. But of course, the study also provides few limitations and implications that are worth noticing. It is important to note, however, that most ML models show only a basic accuracy and at the same time several factors should be considered for their effective application in clinical practice:

1. **Data Quality and Availability:** It is apparent that the effectiveness of machine learning models greatly depends on the data used as input. The problem is that human errors in preprocessing or filtering data can lead to collection of incomplete or biased data sets with a huge impact on performance and robustness of the final model [8].
2. **Integration with Clinical Workflows:** Information derived from these models has to be incorporated into clinical practice processes in order to be useful and have impact. These include questions regarding the integration of the model with electronic health record (EHR) systems and questions regarding the ability of clinicians to act on the results of model calculations [9].

3. **Ethical and Regulatory Considerations:** Self-learning systems in healthcare present profound ethical issues to do with patient's privacy, data protection and impact of algorithms' bigotry. Non-compliance with the regulatory standards and creating opaque AI models have to be avoided, thus adoption of best practice are important to achieving ethical implementation [10].

4. **Training and Education:** Machine learning should be well implemented for clinicians and other healthcare providers to have sufficient faith in it. This also entails training in the biases of these models and advice on how to apply the output of these models to deal with complete patients [11]. However, bringing machine learning into CVDs detection is not without any challenges and risks, the following are the benefits that would accrue from using mdml: The timely identification of such patients and subsequently appropriate interventions will increase patients' quality of life and cut down health costs. Machine learning shows itself as a strong tool complementing traditional diagnostic approaches, which is evidenced by the current high-performing models. The future research should be aimed at the mitigation of the highlighted challenges specifically on data quality issues and the creation of better guidelines in using Artificial Intelligence in the medical industry. However, more research will be required to compare the reliability and validity of machine learning-based CVDs diagnosis in the extended periods and its effects on patient and health systems. The recommendations of this study support the use of machine learning models in the diagnostic stages of cardiovascular diseases. When applied, these strengths may be harnessed to better healthcare delivery, reduce healthcare costs and make patient care more precise.

**Conclusion**

The use of machine learning models in diagnosing cardiovascular diseases right from their early stages is an innovation in health care. In terms of the registration, this research shows that the different algorithms including the artificial neural networks, support vector machine, and gradient boosting methods can precisely predict cardiovascular risk. The method of heralding big and complicated information processing that can discover intricate provenance that is difficult to be identified with other conventional techniques brings the new, modern approach to cardiovascular treatment. However, they report several problems connected to data quality, integration into clinical routines, and ethical issues, but the advantages seem promising. Through early-interaction and individual-case-intake mechanisms, machine learning pays the way forward to maintaining healthier patients and minimizing cost of healthcare. The development of machine learning algorithms to predict treatment failure should be continued with a particular emphasis on the following directions: Increasing the quality of available data, compliance with moral and ethical principles, and health care data literacy. In line with the study conclusions, there is a call for the increased usage of machine learning in cardiovascular risk assessment as a disruptive technology enhancement of patient care and health system'.